\documentclass{article}
\usepackage{spconf,amsmath,graphicx,hyperref}
\usepackage{multirow}   
\usepackage{makecell}
\usepackage{placeins}
\usepackage{float}
\usepackage{subcaption}
\usepackage{geometry}
\usepackage[table]{xcolor}
\usepackage{colortbl}
\usepackage{pgf} 
\usepackage{amssymb}
\usepackage{tikz}
\usepackage[dvipsnames]{xcolor}
\usetikzlibrary{arrows.meta, positioning, shapes.geometric, fit, decorations.pathreplacing}
\tikzset{
    module/.style={rectangle, rounded corners, draw=black, 
                   minimum width=1.0cm, minimum height=1.8cm, 
                   text centered, font=\small},
    process/.style={rectangle, rounded corners, draw=black, fill=Apricot,
                   minimum width=1.6cm, minimum height=1.0cm, text centered, font=\small},
    cnn/.style={rectangle, rounded corners, draw=black, fill=SpringGreen,
                   minimum width=1.0cm, minimum height=1.5cm, text centered, font=\small},
    transf/.style={rectangle, rounded corners, draw=black, fill=SpringGreen,
                   minimum width=1.0cm, minimum height=1.5cm, text centered, font=\small},
    cnn2/.style={rectangle, rounded corners, draw=black, fill=SpringGreen,
                   minimum width=1.0cm, minimum height=1.0cm, text centered, font=\small},
    transf2/.style={rectangle, rounded corners, draw=black, fill=SpringGreen,
                   minimum width=1.0cm, minimum height=1.0cm, text centered, font=\small},
    proj/.style={rectangle, rounded corners, draw=black, fill=White,
                   minimum width=0.8cm, minimum height=1.6cm, text centered, font=\small},
    loss/.style={rectangle, rounded corners, draw=black, fill=gray!20,
                   minimum width=1.6cm, minimum height=1.0cm, text centered, font=\small},
    arrow/.style={-Latex, thick},
}
\newcommand{\heat}[1]{%
    \begingroup
    \pgfmathsetmacro{\v}{abs(#1)}%
    \ifdim \v pt > 0.95pt \cellcolor{green!75!black}{#1}
    \else\ifdim \v pt > 0.9pt \cellcolor{green!95!black}{#1}%
    \else\ifdim \v pt > 0.8pt \cellcolor{green!50!Yellow}{#1}%
    \else\ifdim \v pt > 0.7pt \cellcolor{yellow!70!green}{#1}%
    \else\ifdim \v pt > 0.5pt \cellcolor{yellow!70!red}{#1}%
    \else\ifdim \v pt > 0.3pt \cellcolor{red!55!yellow}{#1}%
    \else \cellcolor{RedOrange}{#1}%
    \fi\fi\fi\fi\fi\fi
    \endgroup
}

\title{DeePAQ: A Perceptual Audio Quality Metric Based On Foundational Models and Weakly Supervised Learning}
\name{
  Guanxin Jiang$^{1}$,
  Andreas Brendel$^{2}$\sthanks{Andreas Brendel has been supported by the Free State of Bavaria by the DSgenAI project.},
  Pablo M. Delgado$^{2}$,
  Jürgen Herre$^{1,2}$
}

\address{
  $^{1}$International Audio 
  Laboratories Erlangen 
  \sthanks{A joint institution of the Friedrich-Alexander Universität Erlangen-Nürnberg (FAU) and Fraunhofer IIS }, Germany \\
  $^{2}$Fraunhofer Institute for Integrated Circuits IIS, Erlangen, Germany
}

\begin{document}
\ninept
\maketitle
\begin{abstract}
This paper presents the Deep learning-based Perceptual Audio Quality metric (DeePAQ) for evaluating general audio quality. Our approach leverages metric learning together with the music foundation model MERT, guided by surrogate labels, to construct an embedding space that captures distortion intensity in general audio. To the best of our knowledge, DeePAQ is the first in the general audio quality domain to leverage weakly supervised labels and metric learning for fine-tuning a music foundation model with Low-Rank Adaptation (LoRA), a direction not yet explored by other state-of-the-art methods. We benchmark the proposed model against state-of-the-art objective audio quality metrics across listening tests spanning audio coding and source separation. Results show that our method surpasses existing metrics in detecting coding artifacts and generalizes well to unseen distortions such as source separation, highlighting its robustness and versatility.
\end{abstract}
\section{Introduction}
\label{sec:intro}
Computational methods have been developed to estimate perceived audio quality as a supplement to subjective evaluation, since applying proper listening tests in every codec development stage is time-consuming, costly, and impractical \cite{9388867}. Computational speech quality assessment has been effectively addressed using contrastive learning and triplet loss. In particular, \cite{10448028, ragano2024scoreq, manocha2021noresqa, manocha2021cdpam} are representative approaches in the speech domain that use models trained with triplet losses in an unsupervised or supervised manner. These methods encode signals into ideally content-agnostic, typically lower-dimensional embeddings using a speech foundation model like wav2vec 2.0 \cite{baevski2020wav2vec}. The Euclidean distance between embeddings of test signals and matched/unmatched reference is assumed to reflect the underlying subjective degradation intensity. Inspired by the strong performance of metric learning and foundation models in speech quality assessment, we extend this paradigm to general audio with a special emphasis on coding artifacts, aiming to develop an audio quality metric that relies on a clean, undistorted reference, i.e., either the same recording without distortion (full-reference) or a different clean recording of a similar signal type (non-matching reference). The challenge of creating such a model is two-fold:

1) Subjective ratings for music content under different types of distortion are much more scarce and rarely publicly available compared to speech quality assessment. As a result, researchers often rely on objective quality assessment tools to bridge the gap in the absence of subjective scores to generate pseudo labels. Large language models are used by \cite{fad, wang2025enabling} to generate text descriptions on audio quality as a surrogate for subjective scores. However, the reliability of these alternative tools is not fully explored as to how faithfully they reflect the audio quality perceived by human listeners, potentially introducing noise into the labels.

2) Compared to speech, music signals display far greater variability, characterized by richer harmonic structures, sharper transients from instruments, such as percussion, and even intentional distortions introduced for artistic expression. Moreover, distortions that are matched to or adapted from the signal content, such as perceptual coding artifacts, are particularly challenging to disentangle, especially when compared to signal-invariant degradations like clipping or additive noise. This diversity highlights the need for powerful foundation models trained on large-scale music datasets to advance general audio quality assessment. Existing music foundation models, such as MERT \cite{li2023mert} and CLAP \cite{laionclap2023,htsatke2022}, are primarily optimized for downstream tasks like music information retrieval and genre classification. The question of which embedding best reflects perceptual aspects of music quality is not yet well understood.

State-of-the-art objective audio quality metrics are intrusive, requiring a clean reference signal to evaluate the quality of a degraded signal under test. A thorough evaluation has been conducted in \cite{9388867} and showed that ViSQOL v3 \cite{chinen2020visqol}, PEAQ \cite{ITU-R_1998}, the 2f-model \cite{8937179}, and HAAQI \cite{Kates2016TheHA} achieve the highest aggregated correlation with human judgments across audio coding and source separation. PEAQ extracts a set of mid-level perceptual features, known as Model Output Variables (MOVs), which are then combined by a small neural network to produce the Overall Difference Grade (ODG). The 2f-model leverages two MOVs from PEAQ Basic \cite{8937179}, resulting in an impressive correlation with subjective scores. HAAQI was designed to assess music quality for hearing-aid applications, but by bypassing its built-in hearing loss simulation, it can also be applied to normal-hearing listeners. Only limited work has explored the potential of music foundation models for perceptual audio quality assessment. Fr\'echet Audio Distance (FAD), used to assess embeddings of generative music models, is highly sensitive to test sample size and the choice of reference signals. Consequently, its reliability is limited, as reflected by the weak Pearson correlation with per-song subjective scores \cite{fad}. A robust tool that exploits a music foundation model for perceptual audio quality remains absent. In this work, we employ the pretrained music foundation model MERT and fine-tune it for general audio quality assessment in a weakly supervised manner using a Rank-n-Contrast (RnC) loss \cite{zha2023rank,ragano2024scoreq}, guided by audio triplets chosen based on surrogate labels.

\section{Proposed Method}
\label{sec:method}
\subsection{Music Foundation Model}
\label{ssec:fmodel}
The proposed approach assumes that the distance between embeddings of test and reference signals reflects perceived audio quality. The embedding function $f : X \to Z$ maps audio samples $\boldsymbol{x}_i \in \mathbb{R}^D$ (where $D$ is the sample length) to a quality embedding space $Z$ so that $f(\boldsymbol{x}_i)$ and $f(\boldsymbol{x}_j)$ are close when $\boldsymbol{x}_i$ and $\boldsymbol{x}_j$ perceived with similar quality and far apart otherwise. The large variability of audio signals, intertwined with imperceptible and thus quality-irrelevant features, complicates the task of mapping high-dimensional data into a low-dimensional embedding space in the desired way. Wav2vec has proven effective in speech quality domain \cite{10448028, ragano2024scoreq}. MERT \cite{li2023mert} shares a similar architecture and self-supervised training strategy as wav2vec \cite{baevski2020wav2vec} and extends it to music signals with an acoustic teacher based on a Variational Autoencoder with residual vector quantization and a music teacher trained with a loss in the Constant-Q Transform domain. 

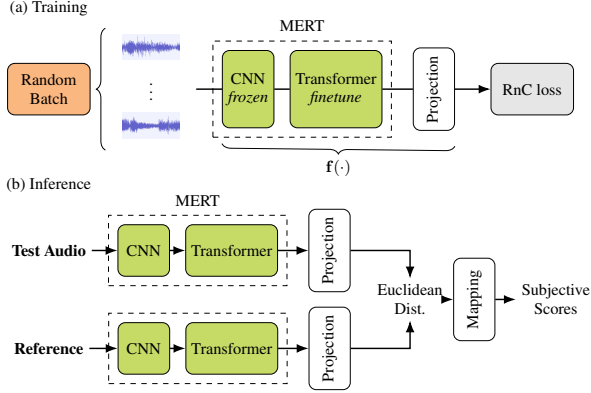
\begin{figure}[t]
\centering
\resizebox{\columnwidth}{!}{%
\begin{tikzpicture}[node distance=0.5cm]
\node[process, align=center] (data) {Random\\Batch};
\node[above=0.8cm of data] {(a) Training};
\node[right=0.3cm of data] (imgs) {
    \begin{tabular}{c}
        \includegraphics[height=0.5cm]{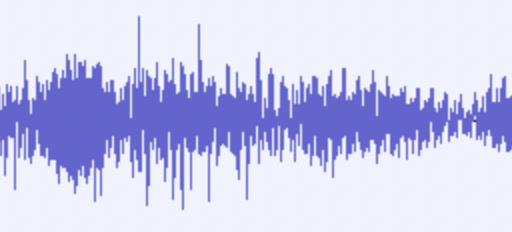} \\[1ex]
        $\vdots$ \\[1ex]  
        \includegraphics[height=0.5cm]{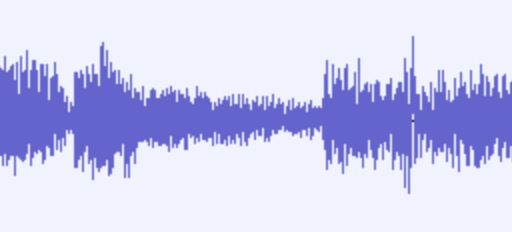}
    \end{tabular}
};
\draw[decorate, decoration={brace, amplitude=5pt, mirror}] ([xshift=-0.01cm]imgs.north west) -- ([xshift=-0.01cm]imgs.south west);
    
\node[cnn, align=center, right=0.5cm of imgs] (cnn) {CNN\\ \textit{frozen}};
\node[transf, align=center, right=0.3cm of cnn] (transf) {Transformer\\ \textit{finetune}};
\node[draw, dashed, inner sep=5pt, fit=(cnn)(transf)] (mertbox) {};
\node[above=2pt of mertbox.north] {MERT};

\node[proj, right=1cm of transf, anchor=center] (proj) {\rotatebox{90}{Projection}};

\draw[decorate, decoration={brace, amplitude=5pt, mirror}] 
    ([yshift=-0.3cm]cnn.south west) -- ([yshift=-0.3cm]proj.south east)
    node[below=5pt, midway, align=center] {$\mathbf{f(\cdot)}$};
    
\node[loss, right=1.5cm of proj, anchor=center] (closs) {RnC loss};

\draw[arrow] (data) (imgs) -- (cnn) -- (transf) -- (proj) -- (closs);

\node [below=2.4cm of data](test) {\textbf{Test Audio}};
\node[above=0.8cm of test] {(b) Inference};
\node[cnn2, right=0.5cm of test] (cnn2) {CNN};
\node[transf2, right=0.3cm of cnn2] (transf2) {Transformer};
\node[draw, dashed, inner sep=5pt, fit=(cnn2)(transf2)] (mertbox2) {};
\node[above=2pt of mertbox2.north] {MERT};
\node[proj, right=1cm of transf2, anchor=center] (proj2) {\rotatebox{90}{Projection}};

\node [below=4.3cm of data](ref) {\textbf{Reference}};
\node[cnn2, right=0.55cm of ref] (cnn3) {CNN};
\node[transf2, right=0.3cm of cnn3] (transf3) {Transformer};
\node[draw, dashed, inner sep=5pt, fit=(cnn3)(transf3)] (mertbox3) {};
\node[proj, right=1cm of transf3, anchor=center] (proj3) {\rotatebox{90}{Projection}};

\node[fit=(proj2)(proj3), inner sep=0pt](midbox) {};
\node[right=0.4cm of midbox, align=center](dist) {Euclidean\\Dist.};

\node[proj, right=0.5cm of dist, anchor=center] (mapping) {\rotatebox{90}{Mapping}};

\node[right=0.4cm of mapping, align=center](sg) {Subjective \\ Scores};

\draw[arrow] (test) -- (cnn2);
\draw[arrow] (cnn2) -- (transf2);
\draw[arrow] (transf2) -- (proj2);

\draw[arrow] (ref) -- (cnn3);
\draw[arrow] (cnn3) -- (transf3);
\draw[arrow] (transf3) -- (proj3);

\draw[arrow] (proj2) -| (dist);
\draw[arrow] (proj3) -| (dist);

\draw[arrow] (dist) -- (mapping);
\draw[arrow] (mapping) -- (sg);

\end{tikzpicture}
}
\caption{Overview of proposed method: (a) Fine-tuning MERT with Rank-n-Contrast loss, (b) Inference with clean reference.}
\label{fig:overview}
\end{figure}

\subsection{Weakly Supervised Training Objective}
\label{ssec:loss}
To put special emphasis on coding artifacts, clean audio signals are coded with AAC, Opus and mp3. The training set comprises these degraded signals along with a small, disjoint subset of the original clean signals, randomly sampled in each batch during training. The bitrate is denoted by $b$, with $b=\infty$ assigned to the clean signals. To establish an audio quality ranking, we use ViSQOL v3 \cite{chinen2020visqol} to compute the Mean Opinion Score (MOS) $v$ of each degraded signal relative to its clean reference, ranging from 1 (very annoying) to 5 (imperceptible). Together with the coding bitrates $b$, which roughly indicate audio quality, these MOS scores serve as surrogate labels. The additional bitrate-based labels are introduced to reduce potential noise and bias from relying on a single annotation source and encourage the model to learn perceptual audio quality from multiple perspectives. For the training dataset $S = \{(\boldsymbol{x}_i, v_i, b_i)\}_{i=1}^{M}$, $\boldsymbol{x}_i \in X$ is the waveform of the $i$-th audio sample (clean or coded), $v_i$ is the corresponding ViSQOL surrogate label, $b_i$ is the coding bitrate and $M$ is the number of samples in the dataset. The set of audio samples is $X$, with subsets of clean and coded signals denoted as $X_{\mathrm{clean}}$, $X_{\mathrm{coded}} = X_{\mathrm{aac}} \cup X_{\mathrm{opus}}\cup X_{\mathrm{mp3}}$, respectively. To capture the continuous nature of audio quality degradation, we apply an RnC loss ($L_\text{RnC}$) \cite{zha2023rank}, which ranks the samples in a batch based on their surrogate labels. The per-sample RnC loss is defined over all $N$ samples in a batch
\begin{equation} 
\mathcal{L}^{p}_{\text{RNC}}(\boldsymbol{x}_i) \!= \!
\frac{-1}{N - 1} \!\sum_{\substack{j=1 \\ j \neq i}}^{N} 
\log \!\frac{\exp\!\left(\|f(\boldsymbol{x}_i)-f(\boldsymbol{x}_j)\|_{2}\right)} {\sum\limits_{\boldsymbol{x}_k \in S^p_{i,j}} \!\exp\!\left(\|f(\boldsymbol{x}_i)-f(\boldsymbol{x}_k)\|_{2}\right)}, 
\label{eq:rnc_loss}
\end{equation}
where $S^p_{i,j} := \{\, \boldsymbol{x}_k\in X \;\mid\; k \neq i,\; |y^p_i-y^p_k| \geq |y^p_i- y^p_j| \,\}$ denotes the set of samples that are of higher ranks than $\boldsymbol{x}_j$ in terms of label distance, given $\boldsymbol{x}_i$ as an anchor. The superscript $p\in\{\mathrm{ViSQOL}, \mathrm{aac}, \mathrm{opus}, \mathrm{mp3}\}$ indicates the label type. When $p = \mathrm{ViSQOL}$, the RnC loss uses the ViSQOL pseudo labels $v_i$ for all batch samples. When $p \in \{\mathrm{aac}, \mathrm{opus}, \mathrm{mp3}\}$, it uses the coding bitrates $b_i$ of the corresponding codec. It is important to note that bitrate only provides a meaningful quality ranking for the same codec, i.e., if $\mathbf{x}_i \in X_p$ with $p\in  \{\mathrm{aac}, \mathrm{opus}, \mathrm{mp3}\}$ we choose
$S^p_{i,j} = \{\, \boldsymbol{x}_k\in X\cup X_p \;\mid\; k \neq i,\; |b^p_i-b^p_k| \geq |b^p_i- b^p_j| \,\}$ in \eqref{eq:rnc_loss}. The overall RnC loss is computed as the batch-wise average of the sample-wise RnC losses
\begin{equation}
\mathcal{L}_{\text{RNC}} \!=
\!\frac{1}{N}
\left(\sum_{\substack{i=1}}^{N}\!\mathcal{L}^{\text{ViSQOL}}_{\text{RNC}}(\mathbf{x}_i) + 
\!\sum_{\substack{\mathbf{x}_i\in X_{\mathrm{coded}}}}\!\mathcal{L}^{p}_{\text{RNC}}(\mathbf{x}_i)\right),
\end{equation}
where $p\in  \{\mathrm{aac}, \mathrm{opus}, \mathrm{mp3}\}$ for the second term.

\subsection{Training Strategy}
\label{ssec:train}
We explored several strategies to adapt MERT to audio quality assessment. First, a projection head was appended on top of a frozen pretrained MERT, but this yielded no substantial improvement compared to other approaches. Next, we fine-tuned the transformer layers, which was prone to overfitting with limited training data, although the effect diminished as the dataset size increased. We also adopted Low-Rank Adaptation (LoRA) \cite{hu2022lora}, a method that updates only low-rank matrices inserted into the frozen pretrained weights, allowing the model to adapt with a small number of trainable parameters. 

\section{Experimental Setup}
\label{sec:experiments}
\subsection{Training Setup}
\label{ssec:trainconfig}
The proposed model uses MERT v1 \cite{li2023mert} with 95M parameters using EnCodec \cite{defossez2022high} as the tokenization approach during pre-training and 12 transformer layers, yielding a $13 \times 768$-dimensional feature matrix per time frame. Averaging over the time dimension and flattening the resulting feature matrix into a one-dimensional vector of length $9,984$ yields the input of the subsequent projection head that is composed of a ReLU activation and a linear layer with $256$-dimensional output. 

We used an internal dataset of approximately 460 hours of CD-quality music recorded at 44.1kHz, encoded with Opus, mp3, and AAC using FFmpeg \cite{ffmpeg}. The raw audio was segmented into 4-second clips and randomly split into disjoint subsets per codec and bitrate. Signals were coded at 16, 32, 48, 64, 80, 96, and 128kbps, yielding a training set of 122 hours of coded audio per codec and 45 hours of clean signals. The validation set comprises 50 hours of music, including 8 hours of clean and 14 hours of coded signals per codec. Training and validation sets do not share the same clean audio but are matched in coding conditions. All signals were resampled to 24kHz to match the pretrained MERT model.

For the proposed full-reference model, we use an initial learning rate of $1 \times 10^{-4}$, decaying exponentially by a factor of 0.99 after 10 epochs without improvement. LoRA matrices are inserted into the query and value projection layers of the attention modules, with a rank 8 and a scaling factor 16. A weight decay of 0.01 and dropout rate of 0.05 are applied to the LoRA parameters. The batch size is 32. For the proposed non-matching reference model, fine-tuning the transformer layers with an initial learning rate of $5 \times 10^{-5}$ yields the best performance, while all other configurations remain identical.
\subsection{Test Sets}
\label{ssec:testset}
The results of nine listening tests were gathered to evaluate the proposed methods, which can be divided into two categories: audio coding and source separation. The IgorC96Multiformat test set \cite{IgorC96} comprises 40 items, primarily music, and was designed to compare Opus, AAC, and Ogg Vorbis at 96 kbps against mp3 at 128 kbps. The Open Dataset of Audio Quality (ODAQ) \cite{torcoli2024odaq} contains 240 audio samples, each rated by 26 listeners, processed by six distortion classes at different quality levels: Low-Pass, Pre-Echoes, Spectral Holes, Tonality Mismatch, Unmasked Noise, and Dialogue Enhancement. The MPEG USAC Verification Tests \cite{ISO2011USAC} include three tests evaluating the Basic Audio Quality (BAQ) of Unified Speech and Audio Coding (USAC) compared with AMR-WB+ and HE-AAC v2 at different bitrates. All three tests use the same 24 excerpts, covering music-only, speech-only, and mixed content, encoded under different conditions. Test 1 (USAC t1) contains mono items at low bitrates (8–24 kbps). Test 2 (USAC t2) and Test 3 (USAC t3) use stereo signals at low (16–24 kbps) and high (32–96 kbps) bitrates, respectively.

We used four subsets from the Subjective Evaluation of Blind Audio Source Separation (SEBASS) dataset \cite{9914724}. These listening tests are PEASS BAQ, SAOC DB, SASSEC, and SiSEC08. In all tests except SAOC, listeners evaluated separated signals submitted to community-based source separation campaigns, identified by system name. The SAOC DB differs in that it investigates the perceived quality of separated sources, subsequently enhanced by the MPEG Spatial Audio Object Coding (SAOC) rendering architecture.
Apart from IgorC96Multiformat, which contains in-domain distortion types but unseen signals for the proposed methods, all other listening tests involve only unseen signals with distortions that are out-of-distribution or out-of-domain.

\section{Evaluation}
\label{sec:evaluation}

\begin{table*}[!t] 
\centering
\scriptsize 
\renewcommand{\arraystretch}{1.2} 
\setlength{\tabcolsep}{2.1pt} 

\begin{tabular}{|c|cc|cc|cc|cc|cc|cc|cc|cc|cc|}
\hline
\multirow{3}{*}{\textbf{Test Sets}} 
& \multicolumn{12}{c|}{\textbf{Full Reference}} 
& \multicolumn{6}{c|}{\textbf{Non-Matching Reference}} \\ \cline{2-19}
& \multicolumn{2}{c|}{PEAQ-ODG} 
& \multicolumn{2}{c|}{HAAQI} 
& \multicolumn{2}{c|}{\makecell{ViSQOL v3}} 
& \multicolumn{2}{c|}{2f}
& \multicolumn{2}{c|}{\makecell{Fine-tune \\wav2vec}} 
& \multicolumn{2}{c|}{\makecell{Proposed}}
& \multicolumn{2}{c|}{\makecell{FAD \\ MERT-v1-95M}} 
& \multicolumn{2}{c|}{\makecell{Fine-tune \\wav2vec 2.0}} 
& \multicolumn{2}{c|}{\makecell{Proposed}} \\ \cline{2-19}
& PCC & SRCC 
& PCC & SRCC 
& PCC & SRCC 
& PCC & SRCC 
& PCC & SRCC 
& PCC & SRCC 
& PCC & SRCC 
& PCC & SRCC 
& PCC & SRCC \\ \hline
IgorC96Multiformat & \heat{0.936} & \heat{0.906} & \heat{0.899} & \heat{0.807} & \heat{0.939} & \heat{0.863} & \heat{0.931} & \heat{0.872} & \heat{0.870} & \heat{0.783} & \heat{0.954} & \heat{0.848} & \heat{-0.016} & \heat{-0.023} & \heat{0.429} & \heat{0.241} & \heat{0.825} & \heat{0.569}\\ \hline
ODAQ-Overall & \heat{0.745} & \heat{0.678} & \heat{0.572} & \heat{0.548} & \heat{0.701} & \heat{0.763} & \heat{0.863} & \heat{0.814} & \heat{0.889} & \heat{0.839} & \heat{0.916} & \heat{0.868} & \heat{-0.131} & \heat{-0.088} & \heat{0.425} & \heat{0.428} & \heat{0.583} & \heat{0.559} \\
Dialogue Enhancement & \heat{0.702} & \heat{0.480} & \heat{0.490} & \heat{0.316} & \heat{0.845} & \heat{0.848} & \heat{0.810} & \heat{0.591} & \heat{0.903} & \heat{0.852} & \heat{0.936} & \heat{0.886} & \heat{0.255} & \heat{0.298} & \heat{0.308} & \heat{0.223} & \heat{0.575} & \heat{0.578}\\
Low-Pass & \heat{0.962} & \heat{0.976} & \heat{0.775} & \heat{0.923} & \heat{0.958} & \heat{0.939} & \heat{0.977} & \heat{0.969} & \heat{0.920} & \heat{0.836} & \heat{0.964} & \heat{0.938} & \heat{-0.167} & \heat{-0.279} & \heat{0.625} & \heat{0.615} & \heat{0.785} & \heat{0.896}\\
Pre-Echoes & \heat{0.880} & \heat{0.847} & \heat{0.615} & \heat{0.560} & \heat{0.687} & \heat{0.920} & \heat{0.962} & \heat{0.975} & \heat{0.961} & \heat{0.938} & \heat{0.966} & \heat{0.941} & \heat{-0.393} & \heat{-0.261} & \heat{0.385} & \heat{0.315} & \heat{0.446} & \heat{0.400}\\
Spectral Holes & \heat{0.693} & \heat{0.514} & \heat{0.685} & \heat{0.612} & \heat{0.485} & \heat{0.579} & \heat{0.941} & \heat{0.927} & \heat{0.949} & \heat{0.848} & \heat{0.874} & \heat{0.797} & \heat{-0.042} & \heat{-0.115} & \heat{0.221} & \heat{0.331} & \heat{0.547} & \heat{0.479}\\
Tonality Mismatch & \heat{0.752} & \heat{0.740} & \heat{0.592} & \heat{0.591} & \heat{0.651} & \heat{0.815} & \heat{0.832} & \heat{0.866} & \heat{0.840} & \heat{0.834} & \heat{0.927} & \heat{0.910} & \heat{-0.034} & \heat{-0.080} & \heat{0.509} & \heat{0.526} & \heat{0.597} & \heat{0.557}\\
Unmasked Noise & \heat{0.850} & \heat{0.938} & \heat{0.668} & \heat{0.703} & \heat{0.675} & \heat{0.807} & \heat{0.867} & \heat{0.901} & \heat{0.790} & \heat{0.801} & \heat{0.896} & \heat{0.895} & \heat{-0.416} & \heat{-0.328} & \heat{0.597} & \heat{0.501} & \heat{0.613} & \heat{0.570} \\ \hline
USAC t1-Overall & \heat{0.532} & \heat{0.422} & \heat{0.433} & \heat{0.429} & \heat{0.893} & \heat{0.895} & \heat{0.857} & \heat{0.874} & \heat{0.804} & \heat{0.774} & \heat{0.900} & \heat{0.877} & \heat{0.009} & \heat{0.031} & \heat{0.478} & \heat{0.435} & \heat{0.673} & \heat{0.623} \\ 
Music & \heat{0.509} & \heat{0.371} & \heat{0.440} & \heat{0.413} & \heat{0.900} & \heat{0.890} & \heat{0.882} & \heat{0.882} & \heat{0.716} & \heat{0.642} & \heat{0.898} & \heat{0.866} & \heat{0.042} & \heat{0.080} & \heat{0.339} & \heat{0.224} & \heat{0.647} & \heat{0.622} \\
Speech & \heat{0.539} & \heat{0.489} & \heat{0.437} & \heat{0.483} & \heat{0.895} & \heat{0.901} & \heat{0.813} & \heat{0.841} & \heat{0.910} & \heat{0.903} & \heat{0.926} & \heat{0.925} & \heat{0.043} & \heat{-0.004} & \heat{0.700} & \heat{0.700} & \heat{0.678} & \heat{0.628} \\
Mix & \heat{0.540} & \heat{0.435} & \heat{0.409} & \heat{0.398} & \heat{0.900} & \heat{0.902} & \heat{0.858} & \heat{0.880} & \heat{0.858} & \heat{0.845} & \heat{0.901} & \heat{0.862} & \heat{-0.066} & \heat{-0.081} & \heat{0.545} & \heat{0.533} & \heat{0.718} & \heat{0.654}\\ \hline
USAC t2-Overall & \heat{0.469} & \heat{0.208} & \heat{0.303} & \heat{0.132} & \heat{0.835} & \heat{0.835} & \heat{0.755} & \heat{0.625} & \heat{0.785} & \heat{0.738} & \heat{0.875} & \heat{0.826} & \heat{0.020} & \heat{0.051} & \heat{0.457} & \heat{0.414} & \heat{0.690} & \heat{0.656} \\ 
Music & \heat{0.404} & \heat{0.041} & \heat{0.239} & \heat{0.053} & \heat{0.860} & \heat{0.854} & \heat{0.726} & \heat{0.458} & \heat{0.716} & \heat{0.602} & \heat{0.871} & \heat{0.774} & \heat{-0.012} & \heat{-0.004} & \heat{0.329} & \heat{0.226} & \heat{0.693} & \heat{0.684} \\
Speech & \heat{0.552} & \heat{0.396} & \heat{0.403} & \heat{0.355} & \heat{0.824} & \heat{0.838} & \heat{0.793} & \heat{0.805} & \heat{0.815} & \heat{0.764} & \heat{0.867} & \heat{0.910} & \heat{0.090} & \heat{0.056} & \heat{0.588} & \heat{0.597} & \heat{0.682} & \heat{0.683} \\
Mix & \heat{0.470} & \heat{0.218} & \heat{0.298} & \heat{0.137} & \heat{0.829} & \heat{0.834} & \heat{0.796} & \heat{0.695} & \heat{0.863} & \heat{0.874} & \heat{0.912} & \heat{0.861} & \heat{0.017} & \heat{0.064} & \heat{0.544} & \heat{0.439} & \heat{0.691} & \heat{0.610} \\ \hline
USAC t3-Overall & \heat{0.624} & \heat{0.692} & \heat{0.515} & \heat{0.618} & \heat{0.863} & \heat{0.898} & \heat{0.884} & \heat{0.922} & \heat{0.818} & \heat{0.850} & \heat{0.928} & \heat{0.938} & \heat{-0.039} & \heat{-0.010} & \heat{0.514} & \heat{0.332} & \heat{0.750} & \heat{0.647} \\ 
Music & \heat{0.524} & \heat{0.550} & \heat{0.481} & \heat{0.549} & \heat{0.858} & \heat{0.871} & \heat{0.888} & \heat{0.922} & \heat{0.743} & \heat{0.780}& \heat{0.939} & \heat{0.948} & \heat{-0.048} & \heat{-0.043} & \heat{0.375} & \heat{0.164} & \heat{0.701} & \heat{0.579} \\
Speech & \heat{0.747} & \heat{0.803} & \heat{0.615} & \heat{0.705} & \heat{0.815} & \heat{0.926} & \heat{0.893} & \heat{0.921} & \heat{0.856} & \heat{0.892} & \heat{0.888} & \heat{0.945} & \heat{0.036} & \heat{-0.051} & \heat{0.637} & \heat{0.412} & \heat{0.762} & \heat{0.665} \\
Mix & \heat{0.666} & \heat{0.752} & \heat{0.497} & \heat{0.618} & \heat{0.894} & \heat{0.933} & \heat{0.902} & \heat{0.943} & \heat{0.903} & \heat{0.907} & \heat{0.946} & \heat{0.928} & \heat{-0.069} & \heat{-0.072} & \heat{0.644} & \heat{0.416} & \heat{0.802} & \heat{0.734}\\ \hline
Source Separation Overall & \heat{0.834} & \heat{0.706} & \heat{0.883} & \heat{0.656} & \heat{0.646} & \heat{0.808} & \heat{0.953} & \heat{0.881} & \heat{0.898} & \heat{0.747} & \heat{0.919} & \heat{0.787} & \heat{0.196} & \heat{0.282} & \heat{0.415} & \heat{0.417} & \heat{0.310} & \heat{0.314}\\
PEASS & \heat{0.754} & \heat{0.313} & \heat{0.758} & \heat{0.155} & \heat{0.468} & \heat{0.531} & \heat{0.898} & \heat{0.624} & \heat{0.845} & \heat{0.420} & \heat{0.859} & \heat{0.467} & \heat{0.177} & \heat{0.127} & \heat{0.339} & \heat{0.356} & \heat{0.374} & \heat{0.409}\\
SAOC & \heat{0.851} & \heat{0.715} & \heat{0.907} & \heat{0.674} & \heat{0.813} & \heat{0.852} & \heat{0.962} & \heat{0.891} & \heat{0.917} & \heat{0.792} & \heat{0.934} & \heat{0.809} & \heat{0.215} & \heat{0.348} & \heat{0.453} & \heat{0.425} & \heat{0.291} & \heat{0.311}\\ 
SASSEC & \heat{0.815} & \heat{0.800} & \heat{0.857} & \heat{0.725} & \heat{0.787} & \heat{0.849} & \heat{0.956} & \heat{0.921} & \heat{0.889} & \heat{0.789} & \heat{0.920} & \heat{0.868} & \heat{0.115} & \heat{0.167} & \heat{0.515} & \heat{0.513} & \heat{0.352} & \heat{0.354}\\
SiSEC08 & \heat{0.875} & \heat{0.763} & \heat{0.920} & \heat{0.775} & \heat{0.784} & \heat{0.876} & \heat{0.948} & \heat{0.899} & \heat{0.927} & \heat{0.817} & \heat{0.948} & \heat{0.829} & \heat{0.210} & \heat{0.319} & \heat{0.363} & \heat{0.371} & \heat{0.248} & \heat{0.246}\\ \hline
\end{tabular}
\caption{Performance comparison (Pearson -PCC- and Spearman rank -SRCC- correlation coefficients between predictor outputs and subjective scores) of proposed full reference and non-matching reference models with other audio quality measurement tools.}
\label{tab:table}
\end{table*}

\begin{figure}[!t]
    \centering
    \includegraphics[width=\linewidth]{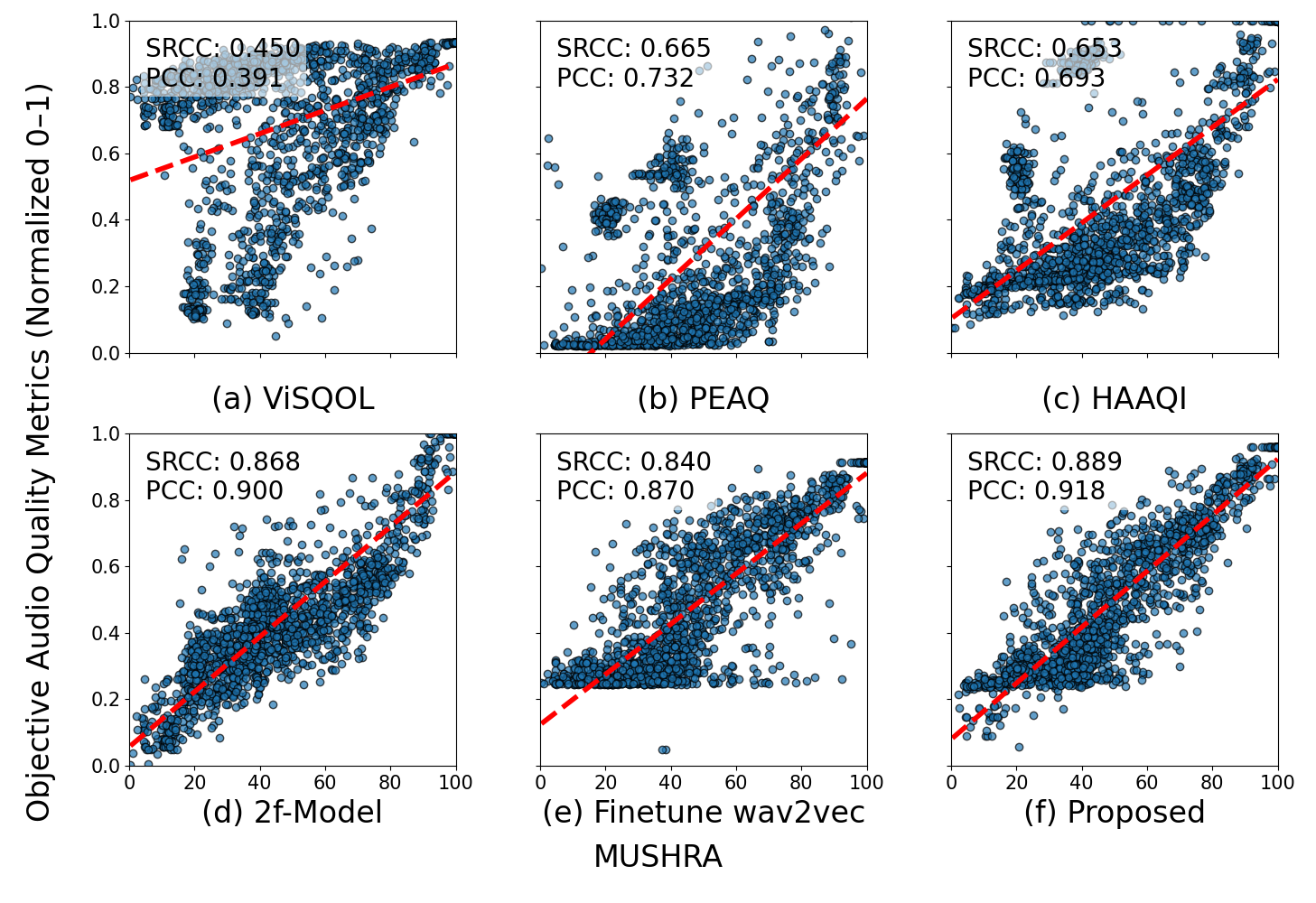}
    \caption{Scatter plots of objective audio quality metric predictions versus subjective scores across all nine listening tests. Subplots (a)–(f) correspond to ViSQOL, PEAQ, HAAQI, the 2f-model, fine-tuned wav2vec 2.0, and the proposed method, respectively. The dashed red line in each subplot shows the linear regression fit to the data points.}
    \label{fig:6subplots}
\end{figure}

\subsection{Baseline Metrics and Results}
\label{ssec:baseline}
To benchmark the proposed models, we incorporate the test results from ViSQOL v3 \cite{chinen2020visqol}, PEAQ ODG \cite{ITU-R_1998}, 2f-model \cite{8937179} and HAAQI \cite{Kates2016TheHA}. We used the MATLAB implementation of the PEAQ Basic version, publicly released by McGill University \cite{kabal2002peaq}. For each listening test, we compute the Pearson linear correlation coefficient (PCC) and the Spearman rank correlation coefficient (SRCC) between predicted and subjective scores. For the proposed methods, the scores are predicted by the cubic polynomial mapping of the Euclidean distance of the embedded test signals relative to the reference embeddings. As an additional baseline, we also finetuned a pretrained wav2vec 2.0 model with the identical setup as for our proposed model. The pretrained BASE wav2vec 2.0 \cite{baevski2020wav2vec} consists of a multi-layer convolutional encoder and 12 transformer layers, similar to MERT-v1-95M. SCOREQ fine-tuned the pretrained BASE wav2vec 2.0 using the SCOREQ loss, which is adapted from the RnC loss \cite{ragano2024scoreq}. Hence, this baseline might be seen as an adaptation of SCOREQ and NOMAD to general audio.

The 2f-model, the fine-tuned wav2vec 2.0, and our proposed full-reference model show the highest overall correlation across all test samples, as shown in Figure~\ref{fig:6subplots}. While the 2f-model excels in the low-quality range, our method shows superior performance in the high-quality range. This may be attributed to the scarcity of training data in the low-quality region. Overall, our method achieves the highest PCC (0.918) and SRCC (0.889). The inclusion of test signals from source separation with out-of-domain distortions reduces the overall performance of our method, yielding a smaller margin over the 2f-model. In Table~\ref{tab:table}, results are illustrated by a background color from red to dark green, representing low to high correlation. Despite strong overall performance of the 2f-model, it struggles with music and mixed items in USAC t2, as well as distortions caused by dialogue enhancement in ODAQ. ViSQOL shows superior performance on USAC t1 and t2 but poor accuracy on ODAQ, particularly for signals with spectral holes. In contrast, the proposed full-reference method demonstrates both high correlation and consistent performance across most test sets, with the exception of PEASS. Interestingly, PEASS proves challenging for all objective metrics evaluated, with only the 2f-model achieving acceptable performance in terms of PCC. The consistently high correlations across both in-domain and out-of-domain tests highlight the robust generalization capability of the proposed model.

The proposed non-matching reference model shows higher effectiveness on audio coding tasks than on source separation, with notably good performance on USAC test sets, where it surpasses PEAQ and HAAQI. We further benchmark it against two non-matching reference baselines, all using the same reference set of 69 clean signals spanning music and speech. The proposed method delivers a marked performance improvement on audio coding compared to the FAD computed on embeddings predicted by the original MERT-v1-95M. 

\subsection{Ablation Study}
\label{ssec:ablation}
An ablation study assessed the impact of the selected foundation model, mapping function, training strategy, and loss. 

\noindent\textbf{Training strategy:} The adaptation techniques for the foundation model were applied to MERT and wav2vec 2.0 under identical settings. In both cases, LoRA achieved the best results by mitigating overfitting on small training datasets, while requiring only 2.93\% of model parameters to be trainable. As the training dataset grew, the performance gap between LoRA and full fine-tuning gradually narrowed. Various configurations of LoRA and transformer layers tuning were explored, including rank sizes, projection layer choices, and learning rate strategies. Among these, the proposed setup achieved the best overall performance on the test sets.

\noindent\textbf{Foundation model:} Predictions from the fine-tuned wav2vec 2.0 model are biased toward speech, showing higher correlations for speech than for music, whereas the proposed method delivers consistent performance across both domains.

\noindent\textbf{Loss function:} Experiments were also conducted to evaluate the inclusion of the RnC loss term for ranking bitrates as additional surrogate labels. The incorporation of this RnC loss term led to a slight performance gain, with improvements of approximately 1–3\% on the test sets.

\noindent\textbf{Mapping function:} An additional observation is that the Euclidean distance between predicted embeddings by the proposed full-reference model exhibit stronger rank correlations than linear correlations in absolute terms. This likely reflects that distances in the embedding space are not linearly related to subjective scores. To address this, a cubic polynomial and a MultiLayer Perceptron (MLP) were explored to map Euclidean embedding distances to subjective scores like MOS/MUSHRA by minimizing mean square error. The MLP comprised three linear layers interleaved with ReLU and Sigmoid activations. Both approaches substantially increased PCC across all test sets, while SRCC remained largely unaffected.

\section{Conclusion}
\label{sec:conclusion}
This paper presents DeePAQ, a perceptual audio quality metric that fine-tunes the music foundation model MERT with LoRA in a weakly supervised setting. The adapted RnC loss encourages the model to learn a quality-related embedding space using only surrogate labels. The proposed full-reference model achieves consistently strong performance in audio coding and generalizes well to out-of-domain scenarios such as source separation. The non-matching reference variant shows clear potential for assessing coding artifacts, with its performance likely to be improved further when trained on a broader range of distortion types.


\clearpage
\bibliographystyle{IEEEbib}
\bibliography{strings,refs}

\begin{thebibliography}{10}

\bibitem{9388867}
M.~Torcoli, T.~Kastner, and J.~Herre,
\newblock ``Objective measures of perceptual audio quality reviewed: An evaluation of their application domain dependence,''
\newblock {\em IEEE/ACM Transactions on Audio, Speech, and Language Processing}, vol. 29, pp. 1530--1541, 2021.

\bibitem{10448028}
A.~Ragano, J.~Skoglund, and A.~Hines,
\newblock ``{NOMAD}: Unsupervised learning of perceptual embeddings for speech enhancement and non-matching reference audio quality assessment,''
\newblock in {\em IEEE International Conference on Acoustics, Speech and Signal Processing (ICASSP)}, 2024, pp. 1011--1015.

\bibitem{ragano2024scoreq}
A.~Ragano, J.~Skoglund, and A.~Hines,
\newblock ``{SCOREQ}: Speech quality assessment with contrastive regression,''
\newblock in {\em Advances in Neural Information Processing Systems}, 2024, vol.~37, pp. 105702--105729.

\bibitem{manocha2021noresqa}
P.~Manocha, B.~Xu, and A.~Kumar,
\newblock ``{NORESQA}: A framework for speech quality assessment using non-matching references,''
\newblock {\em Advances in Neural Information Processing Systems}, vol. 34, pp. 22363--22378, 2021.

\bibitem{manocha2021cdpam}
P.~Manocha, Z.~Jin, R.~Zhang, and A.~Finkelstein,
\newblock ``{CDPAM}: Contrastive learning for perceptual audio similarity,''
\newblock in {\em IEEE International Conference on Acoustics, Speech and Signal Processing (ICASSP)}, 2021, pp. 196--200.

\bibitem{baevski2020wav2vec}
A.~Baevski, Y.~Zhou, A.~Mohamed, and M.~Auli,
\newblock ``wav2vec 2.0: A framework for self-supervised learning of speech representations,''
\newblock {\em Advances in Neural Information Processing Systems}, vol. 33, pp. 12449--12460, 2020.

\bibitem{fad}
S.~Braun D.~Emmanouilidou A.~Gui, H.~Gamper,
\newblock ``Adapting frechet audio distance for generative music evaluation,''
\newblock in {\em IEEE International Conference on Acoustics, Speech and Signal Processing (ICASSP)}, 2024.

\bibitem{wang2025enabling}
S.~Wang, W.~Yu, Y.~Yang, C.~Tang, Y.~Li, J.~Zhuang, X.~Chen, X.~Tian, J.~Zhang, G.~Sun, et~al.,
\newblock ``Enabling auditory large language models for automatic speech quality evaluation,''
\newblock in {\em IEEE International Conference on Acoustics, Speech and Signal Processing (ICASSP)}, 2025, pp. 1--5.

\bibitem{li2023mert}
Y.~Li, R.~Yuan, G.~Zhang, Y.~Ma, X.~Chen, H.~Yin, C.~Lin, A.~Ragni, E.~Benetos, N.~Gyenge, R.~Dannenberg, R.~Liu, W.~Chen, G.~Xia, Y.~Shi, W.~Huang, Y.~Guo, and J.~Fu,
\newblock ``{MERT}: Acoustic music understanding model with large-scale self-supervised training,''
\newblock {\em arXiv preprint:2306.00107}, 2023,
\newblock \url{https://huggingface.co/m-a-p/MERT-v1-95M}.

\bibitem{laionclap2023}
Y.~Wu, K.~Chen, T.~Zhang, Y.~Hui, T.~Berg-Kirkpatrick, and S.~Dubnov,
\newblock ``Large-scale contrastive language-audio pretraining with feature fusion and keyword-to-caption augmentation,''
\newblock in {\em IEEE International Conference on Acoustics, Speech and Signal Processing (ICASSP)}, 2023.

\bibitem{htsatke2022}
K.~Chen, X.~Du, B.~Zhu, Z.~Ma, T.~Berg-Kirkpatrick, and S.~Dubnov,
\newblock ``{HTS-AT}: A hierarchical token-semantic audio transformer for sound classification and detection,''
\newblock in {\em IEEE International Conference on Acoustics, Speech and Signal Processing (ICASSP)}, 2022.

\bibitem{chinen2020visqol}
M.~Chinen, F.~Lim, J.~Skoglund, N.~Gureev, F.~O'Gorman, and A.~Hines,
\newblock ``{ViSQOL} v3: An open source production ready objective speech and audio metric,''
\newblock in {\em 2020 twelfth international conference on quality of multimedia experience (QoMEX)}, 2020, pp. 1--6,
\newblock \url{https://github.com/google/visqol}.

\bibitem{ITU-R_1998}
{International Telecommunication Union} (ITU),
\newblock ``{Method for objective measurements of perceived audio quality},'' ITU-R Recommendation BS.1387-1, 1998.

\bibitem{8937179}
T.~Kastner and J.~Herre,
\newblock ``An efficient model for estimating subjective quality of separated audio source signals,''
\newblock in {\em 2019 IEEE Workshop on Applications of Signal Processing to Audio and Acoustics (WASPAA)}, 2019, pp. 95--99,
\newblock \url{https://audiolabs-erlangen.de/resources/2019-WASPAA-SEBASS}.

\bibitem{Kates2016TheHA}
J.~Kates and K.~Arehart,
\newblock ``The hearing-aid audio quality index ({HAAQI}),''
\newblock {\em IEEE/ACM Transactions on Audio, Speech, and Language Processing}, vol. 24, pp. 354--365, 2016.

\bibitem{zha2023rank}
K.~Zha, P.~Cao, J.~Son, Y.~Yang, and D.~Katabi,
\newblock ``{Rank-N-Contrast}: Learning continuous representations for regression,''
\newblock in {\em Advances in Neural Information Processing Systems}, 2023.

\bibitem{hu2022lora}
E.~Hu, Y.~Shen, P.~Wallis, Z.~Allen-Zhu, Y.~Li, S.~Wang, L.~Wang, W.~Chen, et~al.,
\newblock ``Lo{RA}: Low-rank adaptation of large language models.,''
\newblock {\em ICLR}, vol. 1, no. 2, pp. 3, 2022.

\bibitem{defossez2022high}
A.~D{\'e}fossez, J.~Copet, G.~Synnaeve, and Y.~Adi,
\newblock ``High fidelity neural audio compression,''
\newblock {\em arXiv preprint arXiv:2210.13438}, 2022.

\bibitem{ffmpeg}
``{FFmpeg: A complete, cross-platform solution to record, convert and stream audio and video},'' \url{https://ffmpeg.org/documentation.html}, 2025,
\newblock Version 7.1.1.

\bibitem{IgorC96}
``Public multiformat listening test,'' \url{https://listening-test.coresv.net/index.htm}.

\bibitem{torcoli2024odaq}
M.~Torcoli, C.~Wu, S.~Dick, P.~Williams, M.~Halimeh, W.~Wolcott, and E.~Habets,
\newblock ``{ODAQ}: Open dataset of audio quality,''
\newblock in {\em IEEE International Conference on Acoustics, Speech and Signal Processing (ICASSP)}, 2024, pp. 836--840.

\bibitem{ISO2011USAC}
{ISO/IEC JTC1/SC29/WG11},
\newblock ``{USAC} verification test report {N12232},''
\newblock Technical report, ISO, 2011,
\newblock [Online]. Available: \url{https://mpeg.chiariglione.org/standards/mpeg-d/unified-speech-and-audio-coding.html}.

\bibitem{9914724}
T.~Kastner and J.~Herre,
\newblock ``The {SEBASS-DB}: A consolidated public data base of listening test results for perceptual evaluation of bss quality measures,''
\newblock in {\em 2022 International Workshop on Acoustic Signal Enhancement (IWAENC)}, pp. 1--5.

\bibitem{kabal2002peaq}
P.~Kabal,
\newblock ``An examination and interpretation of itu-r bs.1387: Perceptual evaluation of audio quality,''
\newblock Technical report, McGill University, 2002,
\newblock Code available at \url{http://www-mmsp.ece.mcgill.ca/Documents/Software/}.

\end{thebibliography}

\end{document}